# Low-energy consumption, free-form capacitive deionisation through nanostructured networks

Cleis Santos,*[†,a,b] Inés V. Rodriguez,[†,b,c] Julio J. Lado,*[,c] María Vila,[b] Enrique García-Quismondo,[c] Marc A. Anderson,[c] Jesús Palma[c] and Juan J. Vilatela[b]

[a] Universität Bremen, Energiespeicher- und Energiewandlersysteme, Bibliothekstraße 1, 28359 Bremen, Germany

[b] IMDEA Materials Institute, Tecnogetafe. Eric Kandel, 2, 28906, Getafe, Madrid, Spain.

[c] Electrochemical Processes Unit, IMDEA Energy Avda. Ramón de la Sagra, 3, 28935, Móstoles, Madrid, Spain

* Tel: ++49/421/2246-7332 cleis.santos@uni-bremen.de, Tel: ++34/91/737 11 20 julio.lado@imdea.org

† These authors contributed equally to this work

**ABSTRACT**

Capacitive Deionization (CDI) is a non-energy intensive water treatment technology. To harness the enormous potential of CDI requires improving performance, while offering industrially feasible solutions. Following this idea, the replacement of costly metallic components has been proposed as a mean of limiting corrosion problems. This work explores the use of nanostructured hybrid networks to enable free-form and metal-free CDI devices. The strategy consists of producing interpenetrated networks of highly conductive flexible carbon nanotube (CNT) fibre fabrics and nanostructured metal oxides, $\gamma Al_2O_3$ and $TiO_2$, through ultrasound-assisted nanoparticle infiltration and sintering. In the resulting hybrids, a uniform distribution of porous metal oxide is firmly attached to the nanocarbon network while the flexibility, high conductivity and low-dimensional properties of the CNTs are preserved. These electrodes present a high porosity (105-118 $m^2 \cdot g^{-1}$), notably low electrical (< 0,1 $k\Omega \cdot cm^2$) and low charge transfer resistance (4 $\Omega$), thus enabling the infiltration of aqueous electrolytes and serving as current collector. In this work we built a large asymmetrical cylindrical CDI device solely made of these electrodes and conventional plastics. The cell provides, high average salt adsorption rates of 1.16 $mg \cdot g^{-1}_{AM} \cdot min^{-1}$ (0.23 $mg \cdot g^{-1}_{CDIunit} \cdot min^{-1}$), low energy consumption (0.18 $Wh \cdot g_{salt}^{-1}$) and stable electrochemical performance above 50 cycles for brackish water desalination.

1. **Introduction**

The alarming worldwide problems associated with the scarcity of freshwater are only expected to grow in the coming decades as a consequence of an increasing population and a lack of action to combat environmental concerns.[1] Capacitive Deionization (CDI) is a promising desalination technology due to its low energy consumption and high efficiency.[2] CDI is an electrochemical water treatment technology that removes salt ions and/or charged ions by electrostatic adsorption using porous electrodes operating under an electrostatic potential. This water treatment technology is performed in two steps. During charging, ions are absorbed on the electrode surfaces enabling energy storage in the electrical double layer (EDL), following a process analogous to that of regular supercapacitors. In a second step, the system discharges releasing the ions into another stream (brine). By reversing the current/cell voltage or short-circuiting the cell, part of the energy stored in the EDL can be recovered during discharge and used for subsequent CDI cycles, thus reducing the energy consumption of this water treatment process.[1,3–6]

Efficiency in CDI processes requires electrode materials with high salt adsorption capacity (SAC), high average salt adsorption rates (ASAR), and is conducive to having cells with high round-trip efficiency in a corrosive environment. Nanocarbons play a prominent role in the quest to achieve these properties.[7,8] Their inherently large specific surface area, high electrical conductivity and corrosion resistance makes them attractive electrode materials, particularly when combined with metal oxides (MOx) that increase hydrophilicity[3,9,10] and ion selectivity.[11–14] Much of the challenge has been related to developing methods to process nanocarbons and metal oxides into stable networks that can simultaneously maximise porosity and electrical conductivity for ion removal from aqueous media.[3,15–18] A popular approach is to start with nanocarbons in dispersion and integrate the MOx through wet-chemical methods (hydrothermal, sol-gel, etc), followed by filtering, drying, and other steps before consolidation into fully-formed electrodes.[19] The difficulty lies in forming an effective percolating network with small particles and limited control over the aggregation of the two phases upon solvent removal.[19] Typical CDI properties for these materials are SAC of 5-20 mg·g$^{-1}$, ASAR of 0.5-1.5 mg·g$^{-1}$·min$^{-1}$ and specific energy consumption of 3-6 Wh·g$^{-1}$.[20–22]

In an alternative strategy, a highly-conducting percolated nanocarbon network is first produced and then used as a scaffold for the metal oxide. Separating these stages enables pre-assembly of nanocarbons in electrodes reaching electrical conductivity in the range of metals. Integration of the MOx can then be carried out in a controlled manner to maximise the electrolyte/electrode interface at minimal thickness of the poorly-conducting active phase, increasing efficiency of energy storage and conversion processes.[3,23–25] Applying the atomic layer deposition (ALD) methodology, Feng et al. were able to prepare $TiO_2$/CNT electrodes with a promising performance in terms of SAC, 5.1 mg·g$^{-1}$, ASAR, 0.5-1.5 mg·g$^{-1}$·min$^{-1}$, and energy consumption 3-6 Wh g$^{-1}$.[9] Other groups have favoured the use of fabrics of carbon nanotube fibres (CNTf) because of their combined electrical conductivity approaching copper, high toughness for handling and large electrochemical stability.[30,31] In our previous work, we introduced such a strategy by synthesizing electrodes based on γ$Al_2O_3$/CNTf and $SiO_2$/CNTf, and found a remarkable SAC (6.5 mg·g$^{-1}$) at low net energy consumption (0.3 Wh g$^{-1}$) in an asymmetric CDI device.[3] Here, we introduce a new asymmetric CDI cylindrical cell concept with a much higher capacity (13 mg g$^{-1}$) at low energy consumption (0.18 Wh·g$^{-1}$), largely by using TiO$_2$/CNTf as negative electrode. The TiO$_2$ nanostructured network formed contributes to electrolyte infiltration, ion selectivity (see SEI Fig. S1) [9,28] and other processes beneficial for cell performance.[29,30]

**2. Experimental Section**

**2.1. Fabrication of Electrodes.** The CNTf were synthesized by using the direct spinning chemical vapour deposition method, collected as a unidirectional fabric by winding the material on itself and finally densified.[31] The pristine CNTf fabrics employed here were synthesized under the same conditions as that of our previous work.[32] In summary, the reaction was conducted under a hydrogen atmosphere at 1250°C using the reactants: butanol (source of carbon), ferrocene (iron catalyst) and thiophene (promoter). CNTf, mainly comprised of few-layer multiwall CNTs, were spun from a vertical reactor at a winding rate set at 5 m·min$^{-1}$. Once CNTf were collected, they were infiltrated with TiO$_2$ nanoparticles by using the ultrasound-assisted methodology proposed here **Fig. 1**.a. Firstly, the CNTf fabrics were immersed in a dispersion of TiO$_2$ nanoparticles (**Fig. 1**.b). Commercial TiO$_2$ P25 Aeroxide® was chosen as metal oxide source. An ultrasonication bath was used to deposit the TiO$_2$ particles decorating the fibres with this metal oxide (**Fig. 1**.c). The last step of the synthesis was the calcination process (350°C for 2.5 hours with a temperature rate of 2°C min$^{-1}$ in air atmosphere), in which the nanoparticles of TiO$_2$ sintered over the fibres obtaining a flexible hybrid electrode of ~40 cm$^2$ (**Fig. 1**.d). Different TiO$_2$ concentrations (0.05 - 0.5 - 5 g·L$^{-1}$) and ultrasonication times (1 - 10 - 30 min) were tested. γAl$_2$O$_3$-based positive electrodes were also synthesised following the same ultrasound induced synthesis route as that of TiO$_2$. In this case, hybrid electrodes with a ~20%wt. containing γAl$_2$O$_3$ were prepared, based on optimal electrochemical performance found in a previous study.[3]

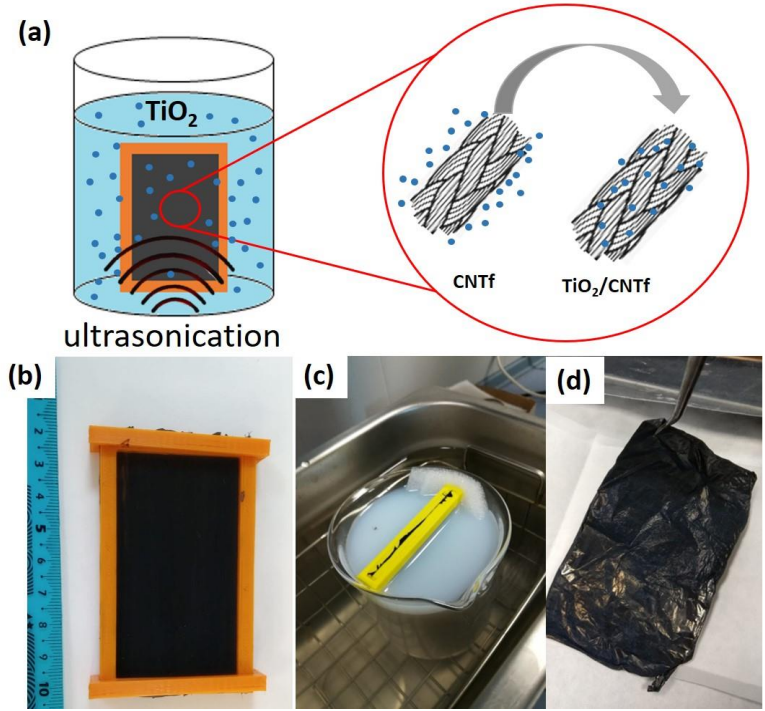

*Fig. 1. Ultrasound-assisted fabrication procedure, a) Scheme of the ultrasonication route for the CNTf impregnation with TiO$_2$ nanoparticles. b) 3D printed frame to comprise the CNT fibre fabric ~40 cm$^2$ c) TiO$_2$-solution with the CNTf mat + frame immersed d) hybrid TiO$_2$/CNTf electrode after the heat treatment step.*

**2.2. Structural & Chemical Characterization.** Structural and chemical characterization was performed by using different techniques. Electron micrographs were obtained by FIB-FEGSEM Helios NanoLab 600i (FEI) at 5 kV. High resolution transmission electron microscopy (HRTEM) images and associated energy-dispersive X-ray spectroscopy (EDS) elemental maps were taken in a JEOL JEM 2100 used at 80 kV. Samples for HRTEM observation were prepared including short sonication for deposition of the material onto TEM grids. These techniques allowed analysis of the network and morphology of the TiO$_2$/CNTf samples. To obtain information about the vibrational modes of the material arising from their interaction with light (photons) via inelastic scattering, Raman spectra (Renishaw) was used, with an excitation wavelength of 532 nm. The content of metal oxide and phase transition was evaluated by Thermogravimetric analysis, TGA, (TA Instruments Q50) at 10°C·min$^{-1}$ in air. X-ray diffraction (XRD) was recorded, using Cu Kα radiation, in Empyrean, PANalytical diffractometer. By using this technique, the interaction of the atomic planes and the crystal lattice was evaluated. Quantachrome instrument (Quadrasorb SI, version 5.03) was employed to measure the nitrogen absorption at 77 K of the samples. Quadrachrome ASiQwin$^{TM}$ software was employed to determine the Specific Surface Area (SSA) by applying the Brunauer–Emmett–Teller (BET) model to the isotherm data points of the adsorption branch in the relative pressure range $P/P_0 < 0.3$. Pore Size Distribution (PSD) was determined from the desorption branch of the isotherms applying the Barrett-Joyner-Halendar (BJH) method.

**2.3. Electrochemical Characterization of Electrodes.** The electrochemical characterization of the sample was performed with an electrochemical workstation (Biologic VMP3 multichannel potentiostat-galvanostat coupled with EC-Lab v11.33 software). Characterizations were conducted in a neutral electrolyte (0.5 M $K_2SO_4$) and using a cellulosic separator between working and counter electrodes. The sample size for these measurements correspond to a circular shape electrode of 0.785 cm$^2$. The three-electrode configuration cell consisted of a $TiO_2$/CNTf as working electrode (WE), platinum mesh as counter electrode (CE) and Ag/AgCl as reference electrode. Under this configuration, cyclic voltammetry (CV), electrochemical impedance spectroscopy (EIS) experiments were performed. To calculate the specific capacitance, cyclic voltammetry experiments were carried out at different scan rates (5 mV s$^{-1}$ to 100 mV s$^{-1}$) establishing a potential window between -0.2 V and 0.8 V. The specific capacitance was calculated using Equation (1) (see ESI). EIS experiments were performed employing a frequency range from 200 kHz to 10 mHz at 0 V. Additionally, V-plots of pristine and hybrid electrodes were obtained by EIS measurements (i.e. at different bias voltage conditions: -0.4 V to 0.8 V) following previous methodology described by Senokos et. al and Santos et al.[23,32] Furthermore, the two-electrode connection cell consisted of $TiO_2$/CNTf as working and counter electrodes which allowed the analysis of the cell performance by galvanostatic charge-discharge measurements. In these experiments, a constant current density of 1 A g$^{-1}$ was applied establishing a maximum cell voltage of 1 V. Cyclability tests were also performed in this two-electrode system by applying the same current density.

**2.4. Cylindrical CDI Device.** An asymmetric cylindrical CDI (ACCDI) cell was assembled using TiO$_2$/CNTf and γAl$_2$O$_3$/CNTf as negative and positive electrodes, respectively, to evaluate the performance of the cylindrical flow-cell for brackish water desalination. The CDI cell takes advantage of the lack of a current collector and flexibility of the CNT fibres. Hybrid electrodes presented ~40 cm$^2$ projected surface area and 7 mg·cm$^{-2}$ mass loading. In this cell, the hybrid electrodes were synthesized by using commercial CNTf, supplied by Tortech Nanofibers. In the ACCDI cell assembling process, the fabricated electrodes were stacked and rolled in a cylindrical configuration. A cellulosic separator was placed between the positive and negative electrodes. To avoid all metallic elements and their associated corrosion issues, CNTf wires were directly connected to the terminals of an electrochemical workstation. Deionization experiments were performed in a batch mode at constant current (max. cell voltage was 1.2 V) using 10 mM NaCl as electrolyte. During these experiments, conductivity was measured in situ while pH and DO (dissolved oxygen) were intermittently monitored to detect possible Faradaic reactions. Altogether, more than 50 cycles in total were evaluated. Salt adsorption metrics were calculated following Hawks et al. [33] Results considering the total mass of the desalination unit are also included for comparison (see ESI).

**3. Results and Discussion**

**3.1. Synthesis of Conformal Hybrids.** The structure of the starting fabrics is a porous graphitic network of CNT bundles with a large specific surface area of ~250 m$^2$·g$^{-1}$ and open mesopores (see SEM micrograph in SEI, Fig. S2).[32,34] A natural and scalable route conducive to the synthesis of large-area CNTf/TiO$_2$ hybrids consists in the infiltration of TiO$_2$ nanoparticles into the fabric of CNTf, followed by sintering. The aim is to use the conductive fabric scaffold as support for TiO$_2$, with controlled integration of the inorganic phase to maximise contact with the CNT and the electrolyte. We performed a comprehensive structural and electrochemical analysis of different samples prepared modifying the two dominant parameters in the synthetic route: nanoparticle concentration and ultrasonication-infiltration time. As expected, the mass fraction of TiO$_2$ increased as nanoparticle concentration increased. For samples infiltrated for 30 min, for example, mass fractions were 17%wt., 22%wt., and 31%wt. at increasing concentrations of 0.05 g·L$^{-1}$, 0.5 g·L$^{-1}$ and 5 g·L$^{-1}$, respectively. Similarly, mass fraction increased with prolonged sonication between 1 and 30 minutes (see TG Analysis in SEI, Fig. S3). More importantly, although the TiO$_2$ mass fraction was in a relatively narrow range, the morphology of the hybrid proved overly sensitive to processing conditions. **Fig. 2** shows the electron micrographs of the different morphologies obtained by adjusting concentration and sonication time. The electrode morphology shows a continuous network of crystalline CNTs (see HRTEM image, **Fig. 2a**) forming a conducting backbone supporting nanoparticles of TiO$_2$ attached to it (see in SI the EDS elemental mapping of the structure of the hybrid electrodes Fig. S4). For high sonication times of 30 minutes, a low TiO$_2$ solution concentration (0.05 g·L$^{-1}$) produced scattered particle agglomerates with only partial coverage of the surface of the CNT bundles (**Fig. 2.**b). Higher nanoparticles concentrations (5 g·L$^{-1}$) produced a thick layer overlay, forming a crust on the CNTf (**Fig. 2.**d). An intermediate concentration led to the formation of a thin (≈ 26-48 nm) continuous layer of TiO$_2$ throughout the CNT fibre fabric (**Fig. 2.**c). Accordingly, an intermediate concentration of 0.5 g·L$^{-1}$ was selected to study the effect of sonication time (**Fig. 2.**c,e,f). In the sample prepared with the shortest sonication time, 1 min, TiO$_2$ nanoparticles formed disjointed aggregates throughout the fabric (**Fig. 2.**e). By extending the sonication time, the TiO$_2$ phase formed a continuous, thin layer on the CNTs, as observed for both 10 minute and 30 minutes infiltration times (**Fig. 2.**f,c). Although there is room for improvement, the examples in **Fig. 2.**c,f are fairly optimised in terms of maximising CNT coverage at a minimal thickness of the inorganic coating.

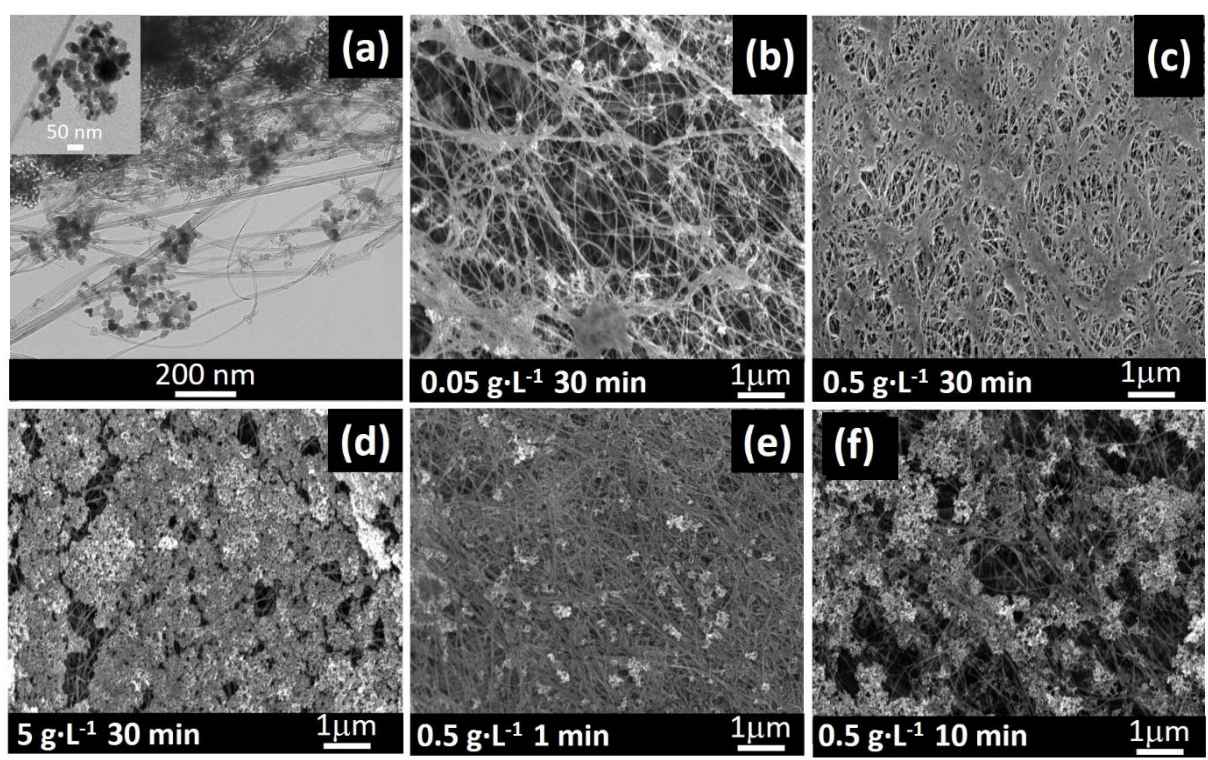

*Fig. 2.* (a) High-resolution TEM (HRTEM) micrographs of sample with a 19%wt. TiO$_2$ showing the network structure of the TiO$_2$ nanoparticles and the CNT fibre. Inset: HRTEM Micrographs of the TiO$_2$/CNTf hybrid electrodes synthesised. Nanoparticle size: 26 ± 5 nm. (b,c,d) for a sonication time period of 30 min at different TiO$_2$ solution concentration 0.05-0.5-5 g·L$^{-1}$ and (d,e,f) using 0.5 g·L$^{-1}$ at different sonication times: 1-10-30 min. The FIB-SEM morphological characterization reveals the variation in the distribution of the nanoparticles, and in the type of hybrid (i.e. particulate and conformal), depending on the synthesis conditions used. (See also SEI Fig. S5).

The influence of the type of morphology (i.e. not well distributed, particulate hybrid, and conformal hybrid) on the porous network, specific surface areas (SSA), and pore distribution were determined by nitrogen gas adsorption. **Table 1** collects the results of metal oxide content, SSA and pore volume of samples synthesized in 0.5 g·L$^{-1}$ at different periods of sonication. Results from the BET and BJH analysis show a clear reduction of the SSA when TiO$_2$ was incorporated to CNTf, regardless the sonication time. This outcome appears to be linked to the formation of an interpenetrating nanostructured network between the metal oxide and the CNTf. A significant reduction of the pore volume (more than 50%) was detected in all the samples, attributed to pore blocking by the small-diameter particles (i.e. 21 nm), as also observed in related hybrids[35,36] Since one of the purposes of hybridising the MOx with CNT is to form a large interface for charge transfer processes,[25] it is expected that the interface forms largely at the expense of a drop in SSA. Overall, these hybrid electrodes preserve a high SSA accessible to water, while minimising the distance between the MOx external surface and the hybrid electrode interface.

*Table 1. Metal oxide content, SSA, pore volume and pore radius of hybrid electrodes synthesized in 0.5g·L$^{-1}$.*

|  | %wt TiO$_2$ | SSA (m$^2$·g$^{-1}$) |  | Pore Volume (cc·g$^{-1}$) | Pore Radius (Å) |
|---|---|---|---|---|---|
| **Electrode** | *TGA* | *BET* | *BJH* | *BJH* | *BJH* |
| pCNTf | - | 250[23,32] | - | 0.89[23] | 152[32] |
| 1min-TiO$_2$/CNTf | 12 | 105 | 115 | 0.29 | 19.0 |
| 10min-TiO$_2$/CNTf | 19 | 113 | 129 | 0.34 | 17.3 |
| 30min-TiO$_2$/CNTf | 22 | 118 | 134 | 0.40 | 17.5 |

Raman and XRD were used to examine the composition for both pristine CNTf and TiO$_2$-decorated samples. This compositional characterization is summarized in **Fig. 3**. For all samples (bare and samples with metal oxide nanoparticles), three characteristic peaks of CNTf can be observed: *1)* D peak, 1350 cm$^{-1}$ ($I_D$ related to defects), *2)* G peak 1580 cm$^{-1}$ ($I_G$, the tangential vibrational mode, i.e. bond stretching of all pairs of sp$^2$ atoms in both rings and chains) and *3)* 2D peak ($I_{2D}$, it is an overtone of G-band thus, related to the vibration of the hexagonal lattice sp$^2$).[31,37–39] **Fig. 3**.a shows an increase of the $I_D/I_G$ ratio with increasing TiO$_2$ content. The introduction of defects occurs by introducing oxidative functional groups on the CNTs during the sintering stage of the hybridization processes. In addition, a peak at 2900-2945 cm$^{-1}$ was also observed in all TiO$_2$-CNTf electrodes. This peak is the D+G overtone peak which is most likely attributed to charge transfer phenomena induced by oxygen-containing functional groups acting as electron withdrawing and accordingly, this peak is also detected in surface functionalized CNTf.[40,41] Since such oxidation takes place only in the presence of TiO$_2$, it is likely indicating a strong coupling of the metal oxide nanoparticles with the carbon nanotubes through a Ti-O-C bond, predicted to be very stable[42] and observed in related nanocarbon/TiO$_2$ hybrids produced thermochemically. Certainly, the CNT/TiO$_2$ interface in the present samples is surprisingly tough, with no evidence of detachment upon extensive handling during testing or assembly into complex-shaped cells (see below and SEI Fig. S6).

The Raman spectra also shows contributions from TiO$_2$ in the range 100-800 cm$^{-1}$ (**Fig. 3**.b). They correspond to anatase [$E_g$ (144 cm$^{-1}$, 196 cm$^{-1}$ and 639 cm$^{-1}$), $B_{1g}$ (396 cm$^{-1}$), 515-519 ($A_{1g}+B_{1g}$)]. No presence of TiO$_2$ in the rutile crystal structure is detected, as confirmed by XRD (Fig. S7). With this assignment, the strong $E_g$ mode for TiO$_2$ can be conveniently used to follow the formation of TiO$_2$ as function of different synthesis parameters (**Fig. 3**.a,b). The data show an expected increase in TiO$_2$ signal with higher

TiO$_2$ concentration or with longer exposure to the dispersion, confirming a larger TiO$_2$ uptake. Interestingly, hybrid samples with higher E$_g$ intensity TiO$_2$ have a more pronounced D' peak and a higher ratio of I$_D$/I$_G$, indicating an increment in the structural defects/disorders in the graphitic structure (inset, **Fig. 3**.c). This trend can be more clearly observed by plotting the ratios of I$_{Eg}$/I$_G$ against I$_D$/I$_G$ (**Fig. 3**.d), which can be taken as "normalised" intensities reflecting the concentration of TiO$_2$ and of defects in the CNTs, respectively. We interpret the positive slope in **Fig. 3**.d as evidence that the presence of TiO$_2$ particles in contact with the graphitic region induces the creation of defects in the CNTs via oxidation during the hybridization process.

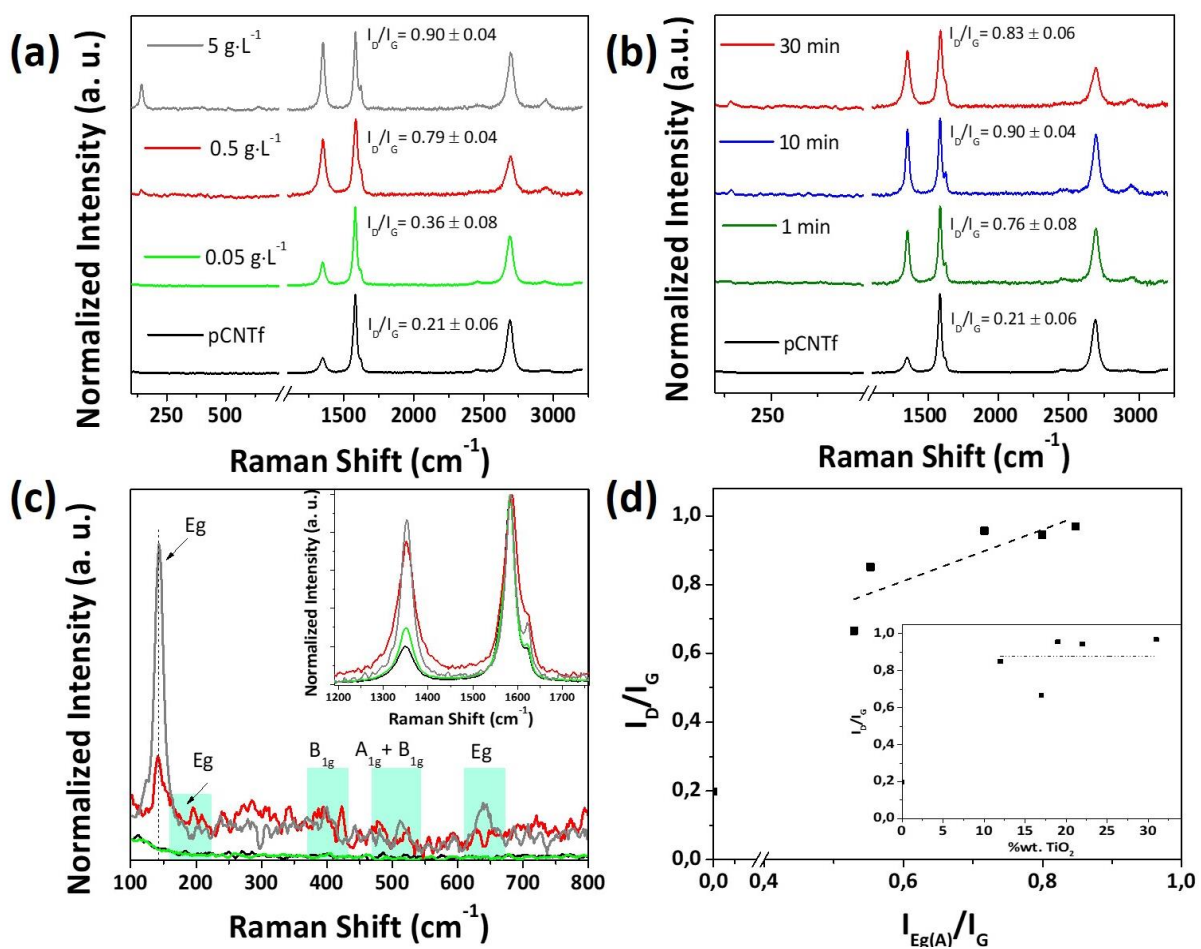

*Fig. 3. Raman analysis of the hybridization process. (a,b) Raman spectra of hybrid electrodes manufactured by sonication during 30 minutes and varying the TiO$_2$ suspension concentration: (a) Extended Raman spectra (b) Raman spectra from 100-800 cm$^{-1}$. In set: modification of the I$_D$/I$_G$ ratio at different solution concentration and the D´ band is more pronounced for the hybrid samples. (c) Extended Raman spectra of*

*hybrid electrodes fabricated by sonication at different periods of time in 0.5 g·L$^{-1}$; (d) Intercorrelation of the defects (I$_D$/I$_G$ ratio) and the interfacial region (I$_{Eg}$/I$_G$ ratio). In set: the I$_D$/I$_G$ ratio remains nearly constant regardless of the TiO$_2$ content in the hybrid electrodes.*

In summary, morphological and spectroscopic analysis shows that the hybrids can be considered as porous networks of nanosized TiO$_2$ anatase crystals uniformly distributed and firmly attached to a continuous conductive network of CNTs.

**3.2. Electrochemical Characterization.** Electrochemical properties of hybrid electrodes were measured to determine optimum hybrid composition for further assembly of a CDI flow cell and in general to gain insight into the role of the two phases in the hybrid. Characterization performed by CV showed a large increase in capacitance as a result of hybridising the CNTf porous structure with TiO$_2$. Upon hybridisation, the voltammograms show an increase in area but otherwise preserve the characteristic "butterfly" shape of the constituent CNTs[23] and absence of faradaic processes involving TiO$_2$ (**Fig. 4.**a). A higher capacitance as compared to the pristine materials was observed for all samples. However, optimum properties were obtained with intermediate TiO$_2$ concentration of 0.5 g L$^{-1}$ in the dispersion and infiltrations > 10 min (**Fig. 4.**b). Samples were also examined under galvanostatic charge-discharge cycling experiments at 23 ± 3 mA·g$^{-1}$, as shown in **Fig. 4.**c for samples prepared using different sonication times. Near optimal concentration, the CD profiles show a triangular line shape and a very small Ohmic drop. These results confirm the trends found by CV of increasing capacity upon hybridisation and the presence of an optimal concentration of TiO$_2$. This optimal composition can be rationalised by analysing the evolution of specific capacitance for different TiO$_2$ contents. **Fig. 4.**d shows that specific capacitance normalised by total electrode mass increases sharply relative to the pristine material, and then remains approximately constant. In contrast, the capacitance normalised by TiO$_2$ content decreases continuously, implying that the contribution to capacitance from TiO$_2$ is most effective at small mass fractions. This effect is in fact caused by the MOx which enabewetting by the electrolyte and infiltration into the otherwise hydrophobic CNTf.[43] The nanostructured $TiO_2$ distributed in the internal pores of the CNTf can drive the ingress of the aqueous electrolyte by providing hydrophilic $TiO_2$ surfaces. Under such a wetting mechanism, $TiO_2$ is discontinuous, but sufficient to enable formation of a large electrolyte/CNTf interface and thus access the effective capacitance of the CNTf material. Previously, we estimated the hydrophobic/hydrophilic transition in CNTf to require a minimum fraction of SiOx nanoparticles of 8 vol.%.[3] All of the hybrids samples in the present work are above such volume fraction.

EIS measurements were also performed on all hybrids. Nyquist plots are shown in **Fig. 4**.e (see equivalent circuit diagram proposed for Fig 4e, SEI Fig. S8). The experimental data could be successfully fitted with a simple equivalent circuit containing a series resistance ($Rs$) connected to an element with a contact resistance ($Rct$) and a capacitance ($C_{tot}$) in parallel.[44] The series resistance represents the resistance from the electrolyte solution in the separator, electrical resistance of current collectors, and resistances of wires. $Rct$ accounts for the interfacial resistance between current collector (i.e. CNTf network), the porous electrodes (i.e. $TiO_2$ nanoparticles) and the charge transfer resistance at the $TiO_2$/CNTf - electrolyte interfaces. Total capacitance ($C_{tot}$) in these samples include the electrosorption Helmholtz capacitance ($C_H$) and the quantum capacitance ($C_Q$), as discussed later. Extracted values obtained from fitting of EIS data are included in Table S1. Values of $Rs$ for all the hybrid samples were low, in the range 0.13 – 0.92 Ω. They confirm that the hybrid has sufficiently high electrical conductivity to be used also as current collector. $Rct$ was found to be lowest for the hybrids produced through infiltration with a medium concentration of 0.5 $g·L^{-1}$, which were also those with the highest capacitance obtained by CV and CD (**Fig. 4**.a-c). These hybrids present a uniform distribution of small nanostructured $TiO_2$ which simultaneously maximise surface area while reducing electrical resistance by preserving a very thin (25 - 40 nm) $TiO_2$/CNTf interconnected network. For these samples, $Rct$ was as low as 4 $\Omega$, which is in fact the value obtained for the pure CNTf material (see ESI **Fig. S9**). As shown in **Fig. 2**.c, using a higher $TiO_2$ concentration during synthesis leads to a large overlayer of $TiO_2$, which translates into larger values of $Rct$. A simple calculation assuming that a hybrid ELDC can be modelled as two series capacitors, one for the CNTf/MOx and for the other MOx/electrolyte interface, shows that for an aqueous electrolyte, capacitance values are high relative to the pristine hydrophobic CNTf, but drop rapidly with increasing metal oxide thickness.[45] EIS measurements also provide more insight into the capacitance dependence on potential, as observed in the CVs in **Fig. 4**.a,b. As discussed elsewhere, such "butterfly shape" voltammograms reflect the contribution from quantum capacitance ($C_Q$) to the total capacitance ($C_{tot}$), in addition to the more common electro-sorption capacitance ($C_H$). The quantum capacitance arises from the low-dimensional electronic structure of the constituent material, namely from a low density of low Density-Of-States (DOS) near the Fermi level.[46,47] Separating these components is challenging because both dependent on the potential drop, and for $C_H$ also varies with electrolyte ionic strength.[48] In **Fig. 4**.f we present total capacitance determined from EIS and normalised by SSA, for different potentials. This representation is convenient because of its resemblance to the line-shape of the DOS; the pristine material (inset of **Fig. 4**.f), for example, shows a V-shape profile consistent with the expected electron-hole symmetry of the joint DOS of CNTs.[23] The first aspect that stands out for the hybrids is the similarity to the profile of the pristine material. This confirms that a significant role of the $TiO_2$ network is in driving the infiltration of aqueous electrolyte and essentially enabling the wetting of the CNTs. However, in the hybrid, there is a branch at negative potential that is significantly steeper and a small feature at -0.1 V. This is attributed to $TiO_2$, which when nanostructured also has a non-negligible quantum (chemical) capacitance due to energy states (i.e. defects) concentrated as an exponential distribution below the conduction band and observed as a capacitance increase with potentials below around 0 V vs Ag/AgCl.[49] This small feature may arise from interfacial states, which, according to simulations on $TiO_2$/graphene hybrids, are located near the conduction band of $TiO_2$.[42] This detailed EIS analysis contributes to our understanding of the electrochemical properties of the hybrid materials synthesised. However, given the small slope in **Fig. 4**.f near the PZC in the potential of interest for CDI, the effect of the electrode's accessible quantum capacitance is minor for water treatment.

From the electrochemical characterisation of hybrid samples, it was concluded that the optimum samples combining high capacitance and low electrical resistance, were those prepared using a $TiO_2$ concentration of 0.5 g L$^{-1}$ and with a concentration of 22 wt.%. It has a specific capacitance value per total mass of electrode (i.e. including both $TiO_2$ and CNTf) of 18 ± 1 F·g$^{-1}$. Such gravimetric capacitance may seem low compared to reported values for higher-surface area EDLC materials tested in aqueous electrolytes and normalised by active material (50-250 F g$^{-1}$).[6,50] However, a more relevant metric for actual CDI cells is the capacity per unit area of electrode and current collector. A comparison with literature data shows that the $TiO_2$/CNTf hybrid in this work outperforms the best reports in the literature by a significant factor (Table S2). Such a comparison highlights the benefits of the present hybrids, where the current collector is effectively built and bound into the electrode and there is no requirement for metallic elements, polymeric binder nor conductive fillers. Equally important is, the high degree of graphitisation of the constituent CNTs in the hybrids responsible for their high electrical conductivity providing high electrochemical stability. A cycle life test was performed at 1 A g$^{-1}$ in a two-electrode cell obtaining an excellent performance without significant decline over 8 000 cycles (ca. 91% capacity retention, see ESI, Fig. S10).

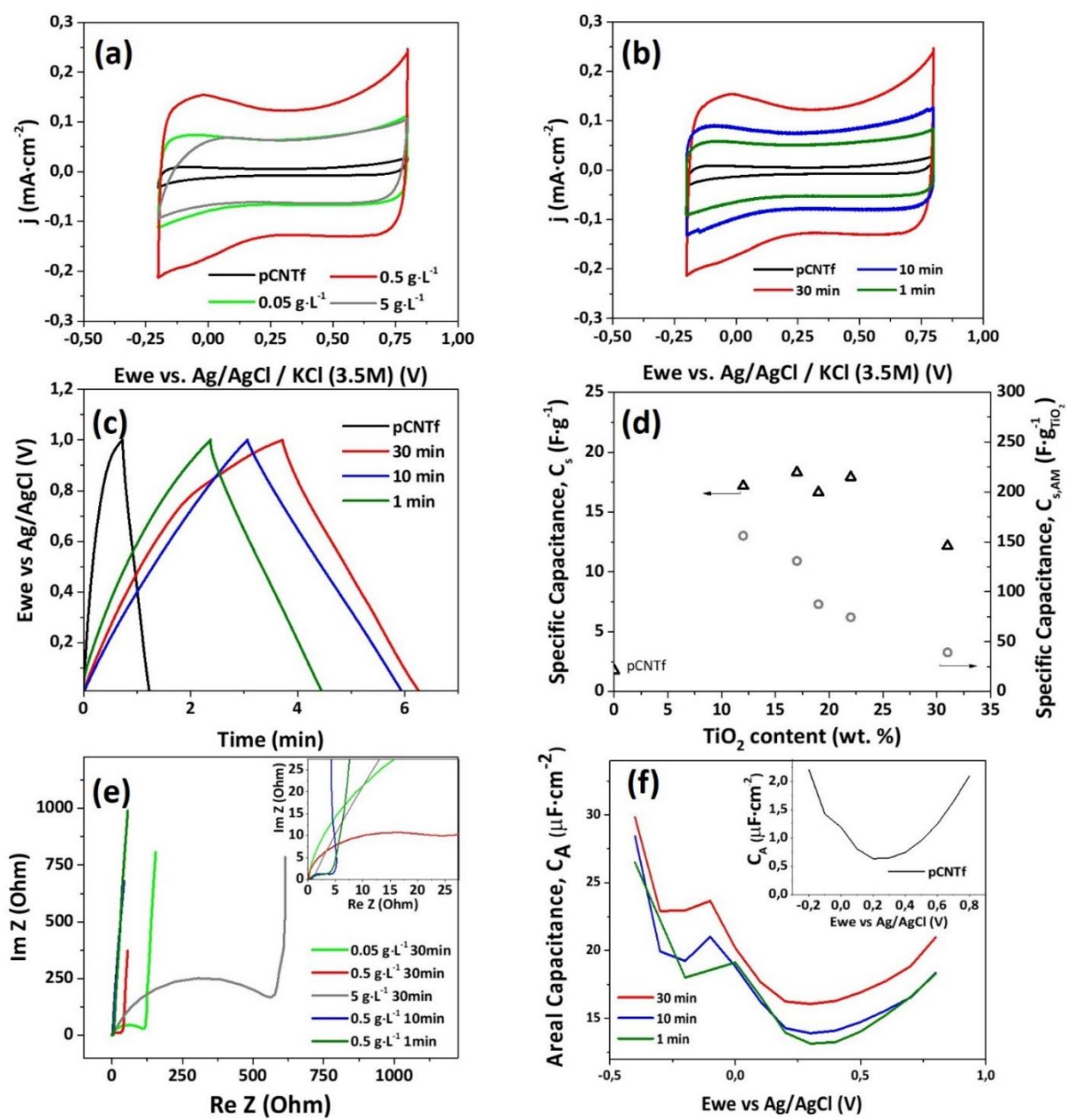

**Fig. 4.** *Electrochemical performance in neutral electrolyte (0.5 M K$_2$SO$_4$). Cyclic voltammetry tests at 20 mV·s$^{-1}$ a) of electrodes synthesized in different TiO$_2$ nanoparticle concentration, b) of electrodes synthesized at different sonication time; c) Galvanostatic charge/discharge measurements at 25 ± 3 mA·g$^{-1}$ of TiO$_2$/CNTf electrodes fabricated at different sonication times and d) Specific capacitance (F·g$^{-1}$) vs. TiO$_2$ content (%wt.); Impedance spectroscopy experiments e) Nyquist plot of the hybrid electrodes. Inset, zoom at high frequencies; f) Area normalized capacitance (C$_A$) vs. Ewe calculated from electrochemical impedance at 10 mHz and different bias voltage of the hybrid electrodes. Inset: C$_A$ vs. Ewe of the pCNTf.*

### 3.3. Metal free asymmetric Cylindrical CDI (ACCDI) cell: proof-of-concept.

Finally, we demonstrate the properties of TiO$_2$/CNTf hybrids as electrodes in a large flow-cell for ACCDI, with a new architecture enabled by their flexibility and elimination of metals by using the hybrids as flexible, low-resistance current collectors. An ACCDI cell was assembled using TiO$_2$/CNTf (*0.5 g·L$^{-1}$ 10 min -TiO$_2$/CNTf*) as the negative electrode and γAl$_2$O$_3$/CNTf as the positive (ACCDI - TiAlOx). Both oxides present opposite superficial charges, negative the TiO$_2$ and positive the Al$_2$O$_3$ (as shown in the Zeta potential as a function of pH determination in Fig. S1) which will potentially favour the electroadsorption of ions, and avoid ions of the opposite charge to be adsorbed during the regeneration step. The γAl$_2$O$_3$/CNTf hybrid electrode was also as reported previously[3] (see SEM images in ESI, Fig. S11).

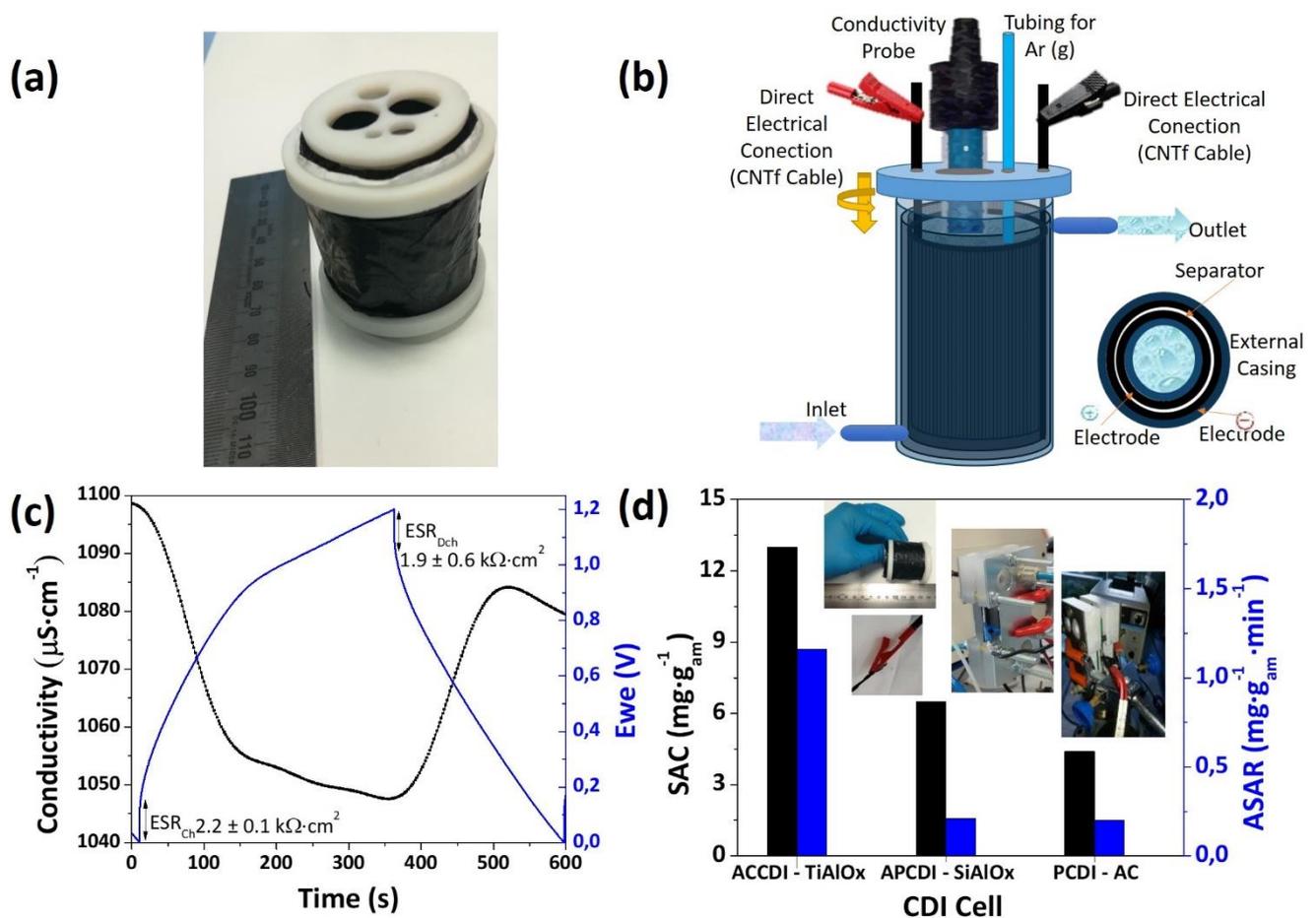

***Fig. 5.*** *Asymmetric cylindrical CDI device. (a) 14 cm$^2$ flexible and metal-free electrodes are tested in the 3D-printed cylindrical cell. (b) Scheme of the cylindrical cell and CNTf*

*cables. Proof-of-Concept of the asymmetric cylindrical CDI device with cables based on CNTf. (c) Ecell and conductivity vs. time profile evolution observed during the proof-of-concept in 10 mM NaCl. (d) Comparison between figures-of-merit of different CDI cell configurations.*

**Fig. 5**.a displays, to the best of our knowledge, the first full cylindrical configuration, without the traditional metallic current collector nor the conventional circular filter-press stacked electrodes configuration. Furthermore, instead of the traditional Cu-based cables to establish the electrical connection, cables made of pristine CNTf were employed (see **Fig. 5**.b). CNTf materials present not only a high SSA but also electrical conductivity in the range of metal and superior mechanical performances (stronger and more flexible).[23,51] They are also chemically stable in electrolytes containing $Cl^-$ without suffering corrosion. The ACCDI cell design was tested using 10 mM NaCl (~600 mg·L$^{-1}$) electrolyte solution. The desalination experiments were performed in constant current mode applying 7.5 mA·g$^{-1}$ (37.5 mA·g$_{AM}^{-1}$) and establishing an operational window between 0 V - 1.2 V. The charge (deionization) – discharge (regeneration) profile displayed in **Fig. 5.**c shows a quasi-triangular profile (Ecell vs. time) indicating successful EDL formation and operation of the metal-free, cylindrical ACCDI. The significant charging-discharging resistances of this cell (2.2 ± 0.1 kΩ·cm$^2$ and 1.9 ± 0.6 kΩ cm$^2$) are probably due to the low conductivity of the electrolyte used (10 mM)[52]. Regarding the two slopes observed in the deionization step, this is attributed to suboptimal balancing of the electrodes in the asymmetric cell. The analysis of the desalination performance revealed that hybrid electrodes assembled in the cylindrical cell showed a high *SAC* of 13 mg·g$^{-1}_{AM}$ (2.5 mg·g$^{-1}_{CDI\ unit}$). This performance represents a large improvement from our previous asymmetric planar CDI (APCDI) $SiO_2$/CNTf - $\gamma Al_2O_3$/CNTf cell and is much higher than a reference control cell consisting of a symmetric planar activated carbon CDI cell (SPCDI – AC) (**Table 2**). As can be noticed in **Fig. 5**.b, the new cylindrical configuration shows an increment in *SAC* by a factor of 3.7 or 5.5 with respect to APCDI-SiAlOx and SPCDI-AC, respectively. Furthermore, the comparison with literature data for other PCDI systems with different electrode materials, shows high performance for the cell in **Fig. 5**.b (**Table 2**). All reported values fall below 13 mg·g$^{-1}$ except for a CDI cells containing reduced graphene oxide (rGO) decorated with TiO$_2$ nanoparticles and tested under different conditions.[21] In fact, recent publications indicate that the maximum *SAC* values obtainable with carbon based capacitive electrodes are in the range of 25 - 30 mg·g$^{-1}$.[53,54], suggesting that the performance of the large cell in this work is approaching practical limits. Moreover, an *ASAR* of 1.16 ± 0.02 mg·g$^{-1}_{AM}$·min$^{-1}$ (0.232 ± 0.004 mg·g$^{-1}_{CDI\ unit}$·min$^{-1}$) was obtained exhibiting superior adsorption rate in comparison with reported results. Thus, the *ASAR* value was significantly higher than those previously tested with PCDI configuration (0.2 mg·g$^{-1}$·min$^{-1}$) for both, APCDI – SiAlOx and PCDI – AC cells. Furthermore, 23.3 L·m$^{-2}$·h$^{-1}$ is the water production result for the ACCDI, considering a salt concentration reduction of 0.3 mM. The charge efficiency was then calculated based on the sequence of charge-discharge profiles along with the conductivity analysis leading to a high value of 85 ± 1 %. Moreover, the calculations determined that an exceptionally low value of specific energy consumption, 0.18 ± 0.01 Wh g$_{salt}^{-1}$ was required by the AC-CDI system. That means that the enhancement in the desalination performance was achieved while preserving a high energy efficiency. Indeed, when comparing with other systems, the net energy consumption was drastically reduced from ~1.9 Wh g$^{-1}$ of the filter-press configuration (i.e. PCDI – AC cell) or from 0.26 Wh·g$^{-1}$ in the case of the APCDI – SiAlOx cell, to 0.18 ± 0.01 Wh·g$^{-1}$. This represents a substantial reduction of 31% of the energy demand and represents

the lowest energy consumption reported for practical CDI process conditions. Moreover, SEM images, TEM observation and Raman measurements confirm that there is no electrode degradation after repeated electrochemical charge/discharge 50 CDI cycles (see ESI, Fig. S12). The electrode morphology shows that the hybrid electrode preserves the continuous CNTf-TiO$_2$. The $I_D/I_G$ ratio shows no variation after cycling and the size of TiO$_2$ particles remain unaltered, therefore implying no electrochemical degradation during deionisation cycles (see Fig. S13). The electrochemical efficiency of the hybrids is inherently very high. In addition, the electric field lines in the cylindrical reactor geometry may contribute to a higher energy efficiency and ion removal compared to a more conventional filter-press reactors.[55,56]

*Table 2. Performance comparison of different CDI systems*

| Electrode Material (Cell Configuration) * | Initial Salt Concentration | SAC** [mg·g$^{-1}$] | ASAR** [mg·g$^{-1}$·min$^{-1}$] | E$_{NET}$ [Wh·g$^{-1}$$_{salt}$] | Cell Voltage [V] | Reference & Year |
|---|---|---|---|---|---|---|
| γAl$_2$O$_3$/CNTf & TiO$_2$/CNTf (ACCDI) | 10 mM | 12.7 ± 0.2 (2.5 ± 0.1) | 1.16 ± 0.02 (0.232 ± 0.004) | 0.18 ± 0.01 | 1.2 | This work |
| γAl$_2$O$_3$/CNTf & SiO$_2$/CNTf (ACCDI) | 34 mM | 6.5 | 0.21 | 0.26 | 1.2 | [3] 2018 |
| Picactif BP10 (SPCDI) | 34 mM | 4.4 | 0.20 | 1.94 | 1.2 | [3] 2018 |
| ALD-TiO$_2$ /CNTs (PCDI) | 86 µS·cm$^{-1}$ | 5.1 | 0.20 | N.A. | 1.6 | [9] 2019 |
| TiO$_2$/AC (MPCDI) | 10 mM | 17.0 | 0.85 | 1.6§ | 1.2 | [29] 2014 |
| TiO$_2$/ACC (PCDI) | 17 mM | 5.3 | 1.55 | 6.60§ | 1.5 | [30] 2018 |
| TiO$_2$/rGO (PCDI) | 150 µS·cm$^{-1}$ | 24.0 | 0.48 | 3.47§ | 1.2 | [57] 2020 |
| TiO$_2$/rGO nanorods (MPCDI) | 5 mM | 16.4 | 1.64 | 3.39§ | 1.2 | [20] 2015 |

*ACCDI, asymmetric cylindrical CDI; ACDI; asymmetric CDI; PCDI, refers to the traditional filter-press electrode stacked (circular or rectangular-shaped electrodes); MPCDI, membrane planar CDI. **Performance CDI metrics considering the total mass of the CDI unit cell (i.e. mass of the CNTf used in the device), noted using parentheses. N.A., Not Available; § Estimated value from the capacitance values and SAC reported in those publications.

## 4. Conclusions

This work introduces a novel cylindrical cell-design approach for CDI based on free-standing flexible TiO$_2$/CNTf electrodes. First, a simple electrode fabrication process based on ultrasound-induced nanoparticle collisions, allowed the preparation of different kinds TiO$_2$/CNTf electrodes (i.e., composites o hybrids). The analysis of the structural and electrochemical properties of the synthesized electrodes showed that a trade-off between the concentration of TiO$_2$ nanoparticles in the suspension and the sonication time is essential to build-up interconnected hybrid networks.

The hybridization process provides better electrochemical performances due to the charge transfer properties of the hybrid material. Based on this evaluation, a conformal hybrid TiO$_2$/CNTf electrode was selected as the best candidate to be used as a negative electrode for the AC-CDI cell. A proof of concept of an ACCDI configuration was then assembled, taking advantage of the flexibility of the CNTf flexible electrodes. Moreover, the high electrical conductivity of CNTf enabled to use a current collector free configuration in which even the wires and terminals were made of CNTf. The ACCDI cell assembled using TiO$_2$/CNTf as the negative electrode and γAl$_2$O$_3$/CNTf as the positive (ACCDI – TiAlOx) showed a great desalination capacity without worsening the energy consumption. The values obtained with the ACCDI could be considered very promising, not only when comparing the previous asymmetrical CDI devices but also to similar TiO$_2$-carbon electrodes, tested under similar conditions. In addition to this, the higher density of the CNTf electrodes allows a reduction of the volume of material required for the electrode preparation.[58] The use of CNTf wiring, which makes the final device 100% metal free, provides additional advantages; such as reducing the final carbon footprint, as well as cutting the final price, due to the increasing cost of metal sources.[59–61] These results make this work an important breakthrough in capacitive deionization technology, with potential for future development of CDI systems with a high ion removal capacity. In this sense, further improvements regarding the optimization of the cylindrical cell design and the development of cables made of CNTf are expected to help outperform the values reported in this manuscript.

**Conflicts of interest**

There are no conflicts to declare.

**Acknowledgements**

Financial support is acknowledged from the European Union Horizon 2020 research and innovation programme under grant agreement 678565 (ERC-STEM), from MINECO (RyC-2014-15115 and HYNANOSC RTI2018-099504-A-C22) and from the Madrid Regional Government (program "Atracción de Talento Investigador", 2017-T2/IND-5568). C. S. acknowledges funding from the European Union's Horizon 2020 research and innovation programme under the Marie Skłodowska-Curie Grant Agreement 84062 (REDEBA). C. S. acknowledges IMDEA Materials funding from the 3rd edition of the Innovation Award Edition 2018. J.J. Lado acknowledges Comunidad de Madrid for the postdoctoral fellowship as part of the Young Talent Attraction Program (2016-T2/AMB-1310). M.V. acknowledges the Madrid Regional Government (program"Atracción de Talento Investigador", 2017-T2/IND-5568) for financial support. The authors also would like to acknowledge Tortech Nanofibers for supplying commercial CNTf fabrics and the technical work performed by Ignacio Almonacid and Guzmán García.

**AUTHOR INFORMATION**

† C. Santos, I. V. Rodriguez, these authors contributed equally to this work

**Corresponding Authors**

*[cleis.santos@uni-bremen.de](mailto:cleis.santos@uni-bremen.de), [julio.lado@imdea.org](mailto:julio.lado@imdea.org)

**REFERENCES**

1   M. A. Shannon, P. W. Bohn, M. Elimelech, J. G. Georgiadis, B. J. Mariñas and A. M. Mayes, *Nature*, 2008, **452**, 301–310.

2   S. Porada, L. Zhang and J. E. Dykstra, *Desalination*, 2020, **488**, 114383.

3   C. Santos, J. J. Lado, E. García-Quismondo, I. V Rodríguez, D. Hospital-Benito, J. Palma, M. A. Anderson and J. J. Vilatela, *J. Mater. Chem. A*, 2018, **6**, 10898–10908.

4   M. A. Anderson, A. L. Cudero and J. Palma, *Electrochim. Acta*, 2010, **55**, 3845–3856.

5   S. Porada, R. Zhao, A. Van Der Wal, V. Presser and P. M. Biesheuvel, *Prog. Mater. Sci.*, 2013, **58**, 1388–1442.

6   Y. Wang, I. Vázquez-rodríguez, C. Santos, E. García-, J. Palma, M. A. Anderson and J. J. Lado, *Chem. Eng. J.*, 2020, **392**, 123698.

7   P. Ratajczak, M. E. Suss, F. Kaasik and F. Béguin, *Energy Storage Mater.*, 2019, **16**, 126–145.

8   J. Han, T. Yan, J. Shen, L. Shi, J. Zhang and D. Zhang, *Environ. Sci. Technol.*, 2019, **53**, 12668–12676.

9   J. Feng, S. Xiong and Y. Wang, *Sep. Purif. Technol.*, 2019, **213**, 70–77.

10  A. S. Yasin, H. Omar, I. M. A. Mohamed, H. M. Mousa and N. A. M. Barakat, *Sep. Purif. Technol.*, 2016, **171**, 34–43.

11  Y. Gao, L. Pan, H. Li, Y. Zhang, Z. Zhang, Y. Chen and Z. Sun, *Thin Solid Films*, 2009, **517**, 1616–1619.

12  H. H. Kyaw, M. Tay, Z. Myint, S. Al-harthi and M. Al-abri, *J. Hazard. Mater.*, 2019, **385**, 121565.

13  S. Kim, H. Yoon, D. Shin, J. Lee and J. Yoon, *J. Colloid Interface Sci.*, 2017, **506**, 644–648.

14  C. Zhao, X. Wang, S. Zhang, N. Sun, H. Zhou, G. Wang, Y. Zhang, H. Zhang and H. Zhao, *Environ. Sci. Water Res. Technol.*, 2020, **6**, 331–340.

15  J. Han, T. Yan, J. Shen, L. Shi, J. Zhang, D. Zhang, P. Iamprasertkun, W. Hirunpinyopas, A. M. Tripathi, M. A. Bissett, R. A. W. Dryfe and J. N. Israelachvili, *Environ. Sci. Technol.*, 2019, **307**, 71–90.

16  G. Divyapriya, K. K. Vijayakumar and I. Nambi, *Desalination*, 2019, **451**, 102–110.

17  A. G. El-Deen, N. A. M. Barakat and H. Y. Kim, *Desalination*, 2014, **344**, 289–298.

18  G. Wang, T. Yan, J. Zhang, L. Shi and D. Zhang, *Environ. Sci. Technol.*, 2020, **54**, 8411–8419.

19  P. Liu, T. Yan, L. Shi, H. S. Park, X. Chen, Z. Zhao and D. Zhang, *J. Mater. Chem. A*, 2017, **5**, 13907–13943.

20  A. G. El-Deen, J. H. Choi, C. S. Kim, K. A. Khalil, A. A. Almajid and N. A. M. Barakat, *Desalination*, 2015, **361**, 53–64.

21 H. M. Moustafa, M. Obaid, M. M. Nassar, M. A. Abdelkareem and M. S. Mahmoud, *Sep. Purif. Technol.*, 2020, **235**, 116178.

22 H. Yin, S. Zhao, J. Wan, H. Tang, L. Chang, L. He, H. Zhao, Y. Gao and Z. Tang, *Adv. Mater.*, 2013, **25**, 6270–6276.

23 E. Senokos, V. Reguero, J. Palma, J. Vilatela and R. Marcilla, *Nanoscale*, 2016, **8**, 3620–3628.

24 A. Pendashteh, J. Palma, M. Anderson, J. J. Vilatela and R. Marcilla, *ACS Appl. Energy Mater.*, 2018, **1**, 2434–2439.

25 M. Rana, V. Sai Avvaru, N. Boaretto, V. A. de la Peña O'Shea, R. Marcilla, V. Etacheri and J. J. Vilatela, *J. Mater. Chem. A*, 2019, 1–17.

26 J. Yang, L. Zou, H. Song and Z. Hao, *Desalination*, 2011, **276**, 199–206.

27 C. Yan, L. Zou and R. Short, *Desalination*, 2012, **290**, 125–129.

28 Y. Cheng, Z. Hao, C. Hao, Y. Deng, X. Li, K. Li and Y. Zhao, 2019, 24401–24419.

29 C. Kim, J. Lee, S. Kim and J. Yoon, *Desalination*, 2014, **342**, 70–74.

30 K. Laxman, D. Kimoto, A. Sahakyan and J. Dutta, *ACS Appl. Mater. Interfaces*, 2018, **10**, 5941–5948.

31 V. Reguero, B. Alemán, B. Mas and J. J. Vilatela, *Chem. Mater.*, 2014, **26**, 3550–3557.

32 C. Santos, E. Senokos, J. C. Fernández-Toribio, Á. Ridruejo, R. Marcilla and J. J. Vilatela, *J. Mat. Chem. A*, 2019, **7**, 5305–5314.

33 S. A. Hawks, A. Ramachandran, S. Porada, P. G. Campbell, M. E. Suss, P. M. Biesheuvel, J. G. Santiago and M. Stadermann, *Water Res.*, 2019, **152**, 126–137.

34 H. Yue, V. Reguero, E. Senokos, B. Mas, R. Marcilla and J. J. Vilatela, *Carbon N. Y.*, 2017, 26.

35  J. J. Wouters, M. I. Tejedor-Tejedor, J. J. Lado, R. Perez-Roa and M. A. Anderson, *J. Electrochem. Soc.*, 2016, **163**, A2733–A2744.

36  J. J. Wouters, M. I. Tejedor-tejedor, J. J. Lado, R. Perez-roa, M. A. Anderson, W. Imdea and E. Alcal, *J. Electrochem. Soc.*, 2018, **165**, 148–161.

37  T. M. G. Mohiuddin, A. Lombardo, R. R. Nair, A. Bonetti, G. Savini, R. Jalil, N. Bonini, D. M. Basko, C. Galiotis, N. Marzari, K. S. Novoselov, A. K. Geim and A. C. Ferrari, *Phys. Rev. B - Condens. Matter Mater. Phys.*, 2009, **79**, 1–8.

38  M. S. Dresselhaus, G. Dresselhaus, R. Saito and A. Jorio, *Phys. Rep.*, 2005, **409**, 47–99.

39  A. C. Ferrari, *Solid State Commun.*, 2007, **143**, 47–57.

40  E. Senokos, M. Rana, C. Santos, R. Marcilla and J. J. Vilatela, *Carbon N. Y.*, 2019, **142**, 599–609.

41  D. Iglesias, E. Senokos, B. Alemán, L. Cabana, C. Navío, R. Marcilla, M. Prato, J. J. Vilatela and S. Marchesan, *ACS Appl. Mater. Interfaces*, 2018, **10**, 5760–5770.

42  L. Ferrighi, G. Fazio and C. Di Valentin, *Adv. Mater. Interfaces*, 2016, **3**, 1500624.

43  A. Pendashteh, E. Senokos, J. Palma, M. Anderson, J. J. Vilatela and R. Marcilla, *J. Power Sources*, 2017, **372**, 64–73.

44  Y. Qu, T. F. Baumann, J. G. Santiago and M. Stadermann, *Environ. Sci. Technol.*, 2015, **49**, 9699–9706.

45  M. Rana, C. Santos, A. Monreal-bernal and J. J. Vilatela, in *Synthesis and Applications of Nanocarbons*, ed. J. A. D. Eder, 2020, pp. 149–200.

46  O. Kimizuka, O. Tanaike, J. Yamashita, T. Hiraoka, S. Saeki, Y. Yamada and H. Hatori, *Carbon N. Y.*, 2008, **46**, 1999–2001.

47  S. Ilani, L. A. K. Donev, M. Kindermann and P. L. McEuen, *Nat. Phys.*, 2006, **2**,

687–691.

48  H. Ji, X. Zhao, Z. Qiao, J. Jung, Y. Zhu, Y. Lu, L. L. Zhang, A. H. MacDonald and R. S. Ruoff, *Nat. Commun.*, 2014, **5**, 1–7.

49  W. H. Leng, P. R. F. Barnes, M. Juozapavicius, B. C. O'Regan and J. R. Durrant, *J. Phys. Chem. Lett.*, 2010, **1**, 967–972.

50  P. Srimuk, M. Zeiger, N. Jäckel, A. Tolosa, B. Krüner, S. Fleischmann, I. Grobelsek, M. Aslan, B. Shvartsev, M. E. Suss and V. Presser, *Electrochim. Acta*, 2017, **224**, 314–328.

51  J. J. Vilatela and R. Marcilla, *Chem. Mater.*, 2015, **27**, 6901–6917.

52  C. Santos, E. García-quismondo, J. Palma, M. A. Anderson and J. Julio, *Electrochim. Acta*, 2019, 135216.

53  F. Yu, L. Wang, Y. Wang, X. Shen, Y. Cheng and J. Ma, *J. Mater. Chem. A*, 2019, 15999–16027.

54  M. E. Suss and V. Presser, *Joule*, 2018, **2**, 10–15.

55  US 2019/0367386 A1, 2019, 1–15.

56  SE 540 976 C2, 2019.

57  H. M. Moustafa, M. Obaid, M. M. Nassar, M. A. Abdelkareem and M. S. Mahmoud, *Sep. Purif. Technol.*, 2020, **235**, 116178.

58  W. Zhang, B. Xu and L. Jiang, *J. Mater. Chem.*, 2010, 6383–6391.

59  Y. Dini, J. Faure-Vincent and J. Dijon, *Carbon N. Y.*, 2019, **144**, 301–311.

60  D. S. Lashmore, *Conductivity mechanisms in CNT yarn*, Elsevier Inc., 2nd edn., 2019.

61  M. Metzger, M. M. Besli, S. Kuppan, S. Hellstrom, S. Kim, E. Sebti, C. V. Subban and J. Christensen, *Energy Environ. Sci.*, 2020, **13**, 1544–1560.